\documentclass[12pt,preprint]{emulateapj}
\usepackage{longtable}
\usepackage{graphicx}
\usepackage{subfigure}
\usepackage{longtable}

\newcommand{\be}{\begin{equation}}
\newcommand{\ee}{\end{equation}}

\newcommand{\msun}{{M}_{\sun}}
\slugcomment{Accepted by ApJ on \today}
\shorttitle{X-ray spectral evolution and radio--X-ray correlation}

\shortauthors{Cao et al.}

\begin{document}
\title{Different X-ray spectral evolution for black hole X-ray binaries in dual tracks of radio--X-ray correlation}
\author{Xiao-Feng Cao\altaffilmark{1}, Qingwen Wu\altaffilmark{1, $\bigstar$}  and Ai-Jun Dong\altaffilmark{1}}

\altaffiltext{1}{School of Physics, Huazhong University of Science and Technology, Wuhan 430074, China}
\altaffiltext{$\bigstar$}{Corresponding author, email: qwwu@hust.edu.cn}

\begin{abstract}
  Recently an `outliers' track of radio--X-ray correlation was found, which is much steeper than the former universal correlation, where dual tracks were speculated to be triggered by different accretion processes. In this work, we test this issue by exploring hard X-ray spectral evolution in four black-hole X-ray binaries (XRBs) with multiple, quasi-simultaneous radio and X-ray observations. Firstly, we find that hard X-ray photon indices, $\Gamma$, are anti- and positively correlated to X-ray fluxes when the X-ray flux, $F_{\rm 3-9keV}$, is below and above a critical flux, $F_{\rm X,crit}$, which are consistent with prediction of advection dominated accretion flow (ADAF) and disk-corona model respectively. Secondly and most importantly, we find that the radio--X-ray correlations are also clearly different when the X-ray fluxes are higher and lower than the critical flux that defined by X-ray spectral evolution. The data points with $F_{\rm 3-9keV}\gtrsim F_{\rm X,crit}$ have a steeper radio--X-ray correlation ($F_{\rm X}\propto F_{\rm R}^{b}$ and $b\sim 1.1-1.4$), which roughly form the `outliers' track. However, the data points with anti-correlation of $\Gamma-F_{\rm 3-9keV}$ either stay in the universal track with $b\sim0.61$ or stay in transition track (from the universal to `outliers' tracks or vice versa). Therefore, our results support that the universal and `outliers' tracks of radio--X-ray correlations are regulated by radiatively inefficient and radiatively efficient accretion model respectively.

\end{abstract}


\section{Introduction}
   Most of Galactic black hole (BH) X-ray binaries(XRBs) are transient sources that are detected in X-rays during bright, several-month-long outbursts with typical recurrence periods of many years. During an outburst, the XRB shows different spectral states associated with a q-like pattern in X-ray hardness-intensity diagram (HID). At the beginning and end of the outburst, XRBs are normally observed in low/hard (LH) state, of which the emission is dominated by a power-law component extending to $\sim$ 100 keV with a photon index of $1.5\lesssim\Gamma\lesssim2$. XRBs will stay in high/soft (HS) state at high luminosities, where the X-ray spectrum is characterized by a strong thermal black-body emission and a weak power-law component with $\Gamma\gtrsim2$. Some XRBs also show a steep power-law state with $\Gamma\gtrsim2.4$, which differs from the HS state in that the power-law component, rather than the disk component, is dominant (\citealt{zd00,mr06,do07,fb12,zs13} for recent reviews).

   Different spectral states in XRBs are believed to be triggered by different accretion and/or jet processes. There is little doubt that a cold, optically thick, geometrically thin standard accretion disk (SSD; \citealt{ss73}) captures the basic physical properties of XRBs in HS state. Soft X-ray bumps as observed in HS state can be naturally explained by multi-temperature blackbody emission of the SSD. The accretion mode for low/hard (LH) state of XRBs is still a matter of debate. The prevalent accretion model is a hot, optically thin, geometrically thick advection-dominated accretion flow (ADAF; also called radiatively inefficient accretion flow, RIAF) that has been developed for BH accreting at low mass accretion rate (e.g., \citealt{nara94,nara95,abra95}; see \citealt{ho08}, \citealt{kato08}, \citealt{na08} and \citealt{yn14} for recent reviews). The low density, two-temperature plasma mainly radiates in X-rays, which can well reproduce the observed spectra of XRBs in LH state. We note that there are also some evidences challenge the pure ADAF model for the LH state of XRBs. The broad Fe K emission line \citep[e.g.,][]{mi06,re08,to08} and/or analysis of disk continuum \citep[e.g.,][]{mi06,re10} suggest that SSD may extend to innermost stable circular orbit in some \emph{bright} hard state of XRBs. For XRBs in a steep power-law state, the accretion process is very unclear, which may be partly caused by the unknown physical mechanism for disk transition (e.g., \citealt{mr06}).

   Numerous studies in the past have shown that hard X-ray photon index, $\Gamma$, is correlated with Eddington ratio for both XRBs and AGNs. \citet{wu08} performed a detailed spectral study for six XRBs in the decay of an outburst and found that $\Gamma$ anti-correlates with $L_{\rm X}/L_{\rm Edd}$ ($L_{\rm Edd}$ is Eddington luminosity) when $L_{\rm X}/L_{\rm Edd}$ is less than a critical value, while they become positively correlated when $L_{\rm X}/L_{\rm Edd}$ is higher than this critical value. This phenomena also exist in AGNs, where $\Gamma$ and Eddington ratio are positively correlated in bright AGNs  \citep[e.g., quasars and bright Seyferts,][]{wang04,sh06,zh10,tr13}, while the anti-correlations are found in a sample of low-luminosity AGNs \citep[e.g.,][]{gc09,con09,yp11}. For a single AGN, the positive correlation of X-ray photon index and luminosity was also found in several bright sources \citep[e.g.,][]{la03,so09}, while the anti-correlation was found in low-luminosity sources \citep[e.g., NGC 7213,][]{em12}. The anti- and positive correlations between $\Gamma$ and Eddington ratio can be regulated by ADAF and SSD/corona respectively as proposed by \citet{wu08}. The detailed model calculations on the ADAF and disk/corona do support this scenario \citep[e.g.,][]{cao09,you12,ql13}. From the X-ray spectral evolution, the LH state can be separated into {\it bright} hard state and {\it dim} hard state, of which the XRBs stay in the positive and anti-correlation of $\Gamma-L_{\rm X}/L_{\rm Edd}$ respectively \citep{wu08}. A small cold disk may exist in some {\it bright} hard state of XRBs due to the condensation of hot accretion flow when accretion rate is larger than $\sim$0.01 Eddington rate, while the inner cold SSD will completely transit to ADAF at lower accretion rate due to the possible evaporation \citep[][]{liu07,liu11}.

   There is a strong connection between radio and X-ray emission in the LH state of XRBs. The quasi-simultaneous radio and X-ray fluxes of more than ten XRBs roughly follow a {\it universal} non-linear correlation of $F_{\rm R}\propto F_{\rm X}^{b}$ ($b\sim0.5-0.7$) as initially found in LH-state GX 339--4 and V404 Cyg \citep[][]{hann98,corb03,gall03}, which seems to be maintained down to quiescent state \citep{gall06}. \citet{wu13} found that the radio--X-ray relation for a sample of AGNs with BH mass of $10^{8\pm0.4}\msun$ is roughly similar to that of LH-state XRBs, where the AGN sample with similar BH mass can be used to simulate the behavior of a single BH XRB in a statistical manner. The quantitative comparison of XRBs and AGNs has made a major step forwards in last ten years with the discovery of a fundamental plane for BH activity in space of BH mass, radio luminosity and X-ray luminosity \citep[e.g.,][]{merl03,falc04,wangr06,ko06,li08,yuan09,gult09,plot12}. The radio spectrum of XRBs in LH state is usually flat or even inverted with spectral index $\alpha\gtrsim0$ ($F_{\nu}\propto\nu^{\alpha}$, $F_{\nu}$ is the radio flux at a certain frequency $\nu$), which is often taken as the evidence for presence of jets \citep[][]{fen01}. The X-ray emission of XRBs is normally thought to originate in the accretion flows \citep[e.g., ADAF,][]{yc05,na08}. Therefore, the tight correlation of radio and X-ray emission suggests that the jet may be coupled with accretion flow. The physical reasons for the jet formation are still unclear, which possibly include a spinning BH \citep[e.g.,][]{bz77,lei05,tc11,li12,nm12}, large-scale magnetic field \citep[e.g.,][]{ca11}, and/or accretion mode \citep[e.g.,][]{me01} etc. The physical reason for the quenched jet in HS state of XRBs is also still unclear, which may be caused by the decrease of the corona due to the stronger cooling when the SSD is present if the jet is coupled with the hot plasma \citep[ADAF/corona,][]{wu13}.

    However, \citet{xc07} found that the radio--X-ray correlation in XRBs might not be as universal as previously thought.  In the following years, more and more XRBs were found to lie well outside the scatter of the former universal radio--X-ray correlation (e.g., XTE J1650$-$500, \citealt{corb04}; XTE J1720-318, \citealt{br05}; Swift 1753.5$-$0127,  \citealt{cado07,sole10}; IGR J17497-2821, \citealt{ro07}; H1743$-$322, \citealt{jonk10,cori11}; XTE J1752$-$223, \citealt{ratt12}). These `outliers' roughly form a much steeper `outliers' track with a correlation slope of $b\sim 1.4$ as initially found in H1743$-$322 \citep{cori11}. The physical reason for the dual tracks of radio--X-ray correlation is still unclear. Assuming the jet launching and radiation behave identically in both tracks, \citet{cori11} speculated that the universal track and `outliers' track may be regulated by radiatively inefficient and radiatively efficient accretion flows respectively based on the several simple scalings of accretion and jet.

    As addressed above, the hard X-ray spectral evolution can shed light on the accretion modes (e.g., ADAF, disk/corona etc.). In this work, we aim to investigate whether the dual tracks of radio--X-ray correlation are regulated by radiatively inefficient and radiatively efficient accretion flows or not, through exploring their hard X-ray spectral evolution. We present the observation and X-ray data reduction in section 2. The results are shown in section 3, which are concluded and discussed in section 4.

\section{Sample selection and X-ray data reduction}
   We consider a sample of BH XRBs in LH state that have multiple, quasi-simultaneous (e.g., within a day) radio and X-ray observations. In order to explore the X-ray spectral evolution for each XRB, we exclude the sources with only one or several observations. The radio flux densities are obtained from literatures directly, while the X-ray data is selected from {\it RXTE} database according to the time of radio observation, and analyzed by ourselves. From these criteria, we select four XRBs (GX 339--4, H 1743--322, Swift J1753.5--0127 and XTE J1752--223). We describe the radio and X-ray data for these sources in more details as follows.

\subsection{Radio data}
   For GX 339--4, we select the radio flux densities at 8.6 or 9.0 GHz from \citet[][Table 1]{corb13}, which were observed by Australia Telescope Compact Array (ATCA) during several outbursts from 1997 to 2012. The 8.5 GHz radio data of H1743--322 are selected from \citet[][and references therein]{cori11}, which were observed by ACTA and Very Large Array (VLA) regularly for several outbursts from 2003 to 2010. For Swift J1753.5--0127, \citet[][]{sole10} presented the radio observation with VLA, Westerbork Synthesis Radio Telescope (WRST) and MERLIN at 1.7--8.5 GHz during an outburst from 2005 to 2009. The radio fluxes of three data points observed at 1.7 GHz by MERLIN and four observed by VLA at 4.8 GHz are converted to fluxes at 8.5 GHz assuming a typical radio spectral index of $\alpha=-0.12$ for LH state of XRBs \citep[$F_{\nu}\propto\nu^{-\alpha}$, e.g.,][]{corb13}. For XTE J1752$-$223, we use the radio flux densities at 9.0 GHz that observed by ATCA during an outburst from 2009 to 2010 \citep[][]{br13}. We list all the radio data in Tables 1-4. In the following analysis, we simply assume the radio flux densities at 8.4-9.0 GHz is more or less similar based on the flat spectrum as observed in most LH-state XRBs, which will not affect our results.

\subsection{X-ray spectral analysis}
   According to the radio data, we select the quasi-simultaneous (within 1 day) X-ray data from {\it RXTE} archive. There are 69 observations for GX 339--4 (Table 1), 28 observations for H 1743--322 (Table 2), 18 observations for Swift J1753.5--0127 (Table 3), and 14 observations for XTE J1752--223 (Table 4). The X-ray data is reduced and analyzed using HEASOFT software following standard steps described in {\it RXTE} cookbook\footnote{\it http://heasarc.gsfc.nasa.gov/docs/xte/data$_{-}$analysis.html}.

   We extract Proportional Counter Array \citep[PCA,][]{ja06} spectra from Standard 2 data, where only top layer of Proportional Counter Unit (PCU) 2 is used since that it is only one operational across all the observations and is the best calibrated detector out of five PCUs. Response matrices are generated and background spectra are created using the latest PCA background model (faint or bright) according to brightness level. We also extract HEXTE \citep[High Energy Timing Experiment,][]{ro99} spectra from archive data, where the data from cluster B is considered. However, the Cluster B was also stopped rocking from December 14th 2009.

   In order to better exploring the hard X-ray spectral evolution, we analyze the X-ray data in the same energy range for a given XRB. Therefore, we perform a simultaneous fit to the PCA (3-25 keV) and HEXTE (20-200 keV) spectra in XSPEC for H 1743-322 and Swift J1753.5-0127, where all the data of these two sources are observed before December 2009. However,  PCA spectra (3-40 keV) are fitted for GX 339--4 and XTE J1752--223 due to many radio and X-ray data are selected from the outbursts after December 2009. We add a systematic uncertainties of 0.5\% to all spectral channels to account for PCA calibration uncertainties. To normalize between PCA and HEXTE, we multiple the HEXTE model by a constant component. The main aim of the spectral analysis is to derive the X-ray spectral index and unabsorbed X-ray flux. Therefore, we use the model as simple as possible. We use a power-law component ({\it powerlaw}) and an absorption component ({\it phabs}) as starting model, where the hydrogen column density was fixed at $N_{\rm H}=0.5\times10^{22} \rm cm^{-2}$ for GX 339--4 \citep[e.g.,][]{ko00}, $N_{\rm H}=1.6\times10^{22}\rm cm^{-2}$ for H 1743--322 \citep[e.g.,][]{cap09}, $N_{\rm H}=1.7\times10^{21} \rm cm^{-2}$ for Swift J1753.5--0127 \citep[e.g.,][]{hi09}, and $N_{\rm H}=0.46\times10^{22} \rm cm^{-2}$ for XTE J1752--223 \citep[e.g.,][]{ma09}. A multi-temperature disk blackbody (diskbb), a Gaussian emission line ({\it gaussian}) and a high energy cutoff ({\it highecut}) will be added if they can improve the fittings substantially (e.g., $\vartriangle\chi^2>0.3$). The broken power-law ({\it bknpower}) will be used to replace the simple power-law component in some spectral fittings if we cannot get satisfactory fitting result with above models (e.g., $\chi^2>2$), where the break energy is around 20 keV (e.g., H 1743--322, \citealt{mc09}, Swift J1753.5-0127, \citealt[e.g.,][]{sole10}. In addition, we subtract the Galactic ridge emission contamination $6.0\times10^{-11} \rm erg s^{-1}cm^{-2}$ \citep[e.g.,][]{cori11} from the unabsorbed flux of H 1743--322 due to the location of H 1743--322 is close to the Galactic plane. The observational date, ID, unabsorbed 3-9 keV X-ray flux, the photon index, fitting results and adopted models are shown in the Tables 1-4.

\section{Results}
%---------------------------------------------Figure 1-----------------------------------------------------------------------------------------------------------------------------------------
\begin{figure}[b]
%\vspace{-0mm}
\epsscale{0.9} \plotone{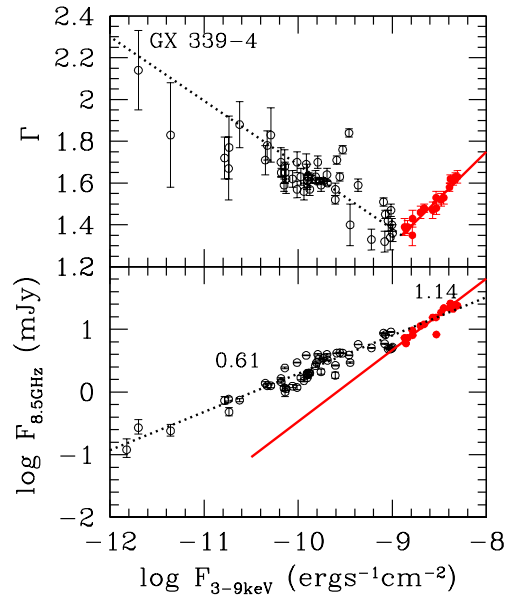}
%\vspace{-8mm}
\caption{Top panel shows the relation between X-ray photon index, $\Gamma$, and unabsorbed 3-9 keV X-ray flux, $F_{\rm 3-9keV}$, for GX 339--4, where we fit the correlation with a piecewise linear regression (two segments). The dotted and solid lines represent the best fits for the data points in anti- (open circles) and positive (solid circles) correlations respectively. The bottom panel shows the radio--X-ray correlations, where the open and solid circles are defined according to their X-ray spectral evolution in the top panel, where the dotted and solid lines are their best fits (correlation slopes are also shown).  \label{fig1}}
\end{figure}

%---------------------------------------------Figure 2-----------------------------------------------------------------------------------------------------------------------------------------
\begin{figure}[b]
%\vspace{-17mm}
\epsscale{0.9} \plotone{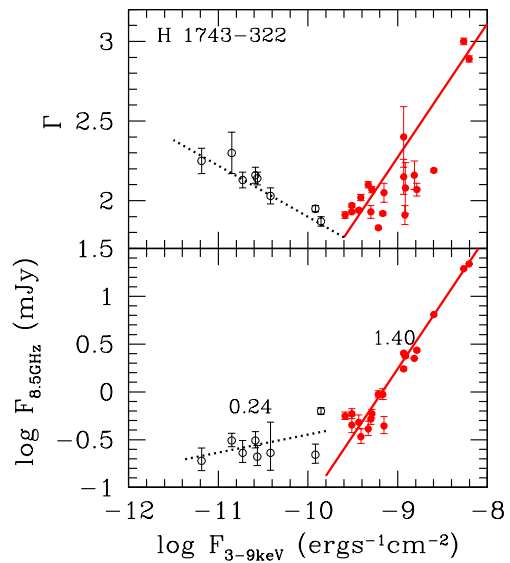}
%\vspace{-12mm}
\caption{The same as Figure 1, but for H 1743-322.  \label{fig1}}
\end{figure}

   We present relations between hard X-ray photon index, $\Gamma$, and 3--9 keV X-ray flux, $F_{\rm 3-9 keV}$, for four sources in top panels of Figures 1-4 respectively, where the photon index at harder band is adopted if it is fitted with a broken power-law model. For GX 339--4 and H 1743--322 (Figures 1 and 2), it can be clearly found that $\Gamma$ is anti-correlated to $F_{\rm 3-9 keV}$ (open circles) when $F_{\rm 3-9 keV}$ is lower than a critical value, $F_{\rm X,crit}$, while these two quantities become positively correlated (solid circles) when $F_{\rm 3-9 keV}\gtrsim F_{\rm X,crit}$. The positive correlation of $\Gamma-F_{\rm 3-9 keV}$ is evident in Swift J1753.5--0127 as a whole, even though the correlation become flat or a little bit anti-correlated when X-ray flux lower than a critical value (see top panel of Figure 3).  For XTE J1752--223, there are some trends in both anti- and positive correlations of $\Gamma-F_{\rm 3-9 keV}$ except that four data points evidently deviate from both anti- and positive correlations (see top panel of Figure 4). We fit the relation of $\Gamma-\log F_{\rm 3-9 keV}$ with a piecewise linear regression (two segments) for each source (top panels in Figures 1-4). Four data points of XTE J1752--223 (with the highest X-ray fluxes) are excluded in the fitting which evidently deviate from both anti- and positive correlations. We list the critical flux (or break point) and its 1 $\sigma$ error for the transition of different X-ray spectral evolutions for each XRB in Table 5, where $F_{\rm X,crit}\simeq1.3\times10^{-9} \rm ergs^{-1}cm^{-2}$ for GX 339--4 is several times higher than that of other two XRBs ($\simeq2.5\times10^{-10} \rm ergs^{-1}cm^{-2}$, H 1743--322 and XTE J1752--223). In the fitting of Swift J1753.5--0127, only upper limit of critical flux is derived ($F_{\rm X,crit}\lesssim2.5\times10^{-10} \rm ergs^{-1}cm^{-2}$ ), which may be caused by the data points in anti-correlation are not evident or too few.

%---------------------------------------------Figure 3-----------------------------------------------------------------------------------------------------------------------------------------
\begin{figure}[t]
%\vspace{-15mm}
\epsscale{0.9} \plotone{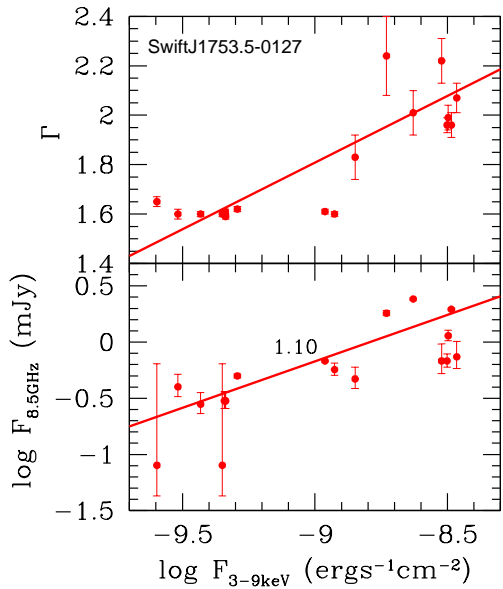}
%\vspace{-12mm}
\caption{The same as Figure 1, but for Swift J1753.5-0127.  \label{fig1}}
\end{figure}

%---------------------------------------------Figure 4-----------------------------------------------------------------------------------------------------------------------------------------
\begin{figure}[t]
%\vspace{-15mm}
\epsscale{0.9} \plotone{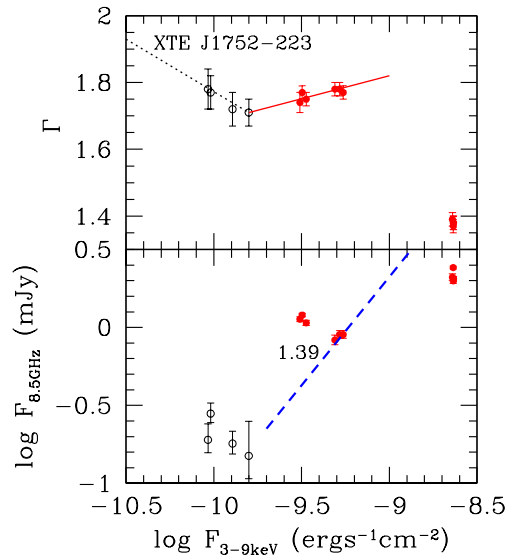}
%\vspace{-12mm}
\caption{The same as Figure 1, but for XTE J1752-223. The dashed line in the bottom panel is not the fitting result for the solid point as defined in the top panel, but the best fit for 17 quasi-simultaneous radio/X-ray data points reported in \citet{br13}, which are observed from several X-ray satellite and all the selected data points have 3-9 keV X-ray higher than the critical flux defined by X-ray spectral evolution of the top panel.   \label{fig1}}
\end{figure}

   We find that the radio--X-ray correlation seems to also change along the change of $\Gamma-F_{\rm 3-9 keV}$ correlation (particularly see Figures 1 and 2). Therefore, we divide the data of each source into one (Swift J1753.5--0127) or two groups (GX 339--4 and H 1743--322) according to the critical X-ray fluxes as defined by $\Gamma-F_{\rm 3-9 keV}$ correlation.  Then we fit the radio--X-ray correlations for each source with a power-law form of $F_{\rm R}=kF_{\rm X}^b$ for the data points with $F_{\rm 3-9 keV}$ lower or higher than $F_{\rm X,crit}$ respectively. In the fittings, the typical flux variations (within one day) $\sigma_{\rm R}=0.1\rm\ dex$ in radio band and $\sigma_{\rm X}=0.15\rm\ dex$ in X-ray band \citep[e.g.,][]{cori11,corb13} are assumed to be their flux uncertainties instead of using the observational flux errors. We list the fitting results in Table 5, where $b_1$ and $b_2$ represent the correlation slopes for the data points with $F_{\rm 3-9 keV}\lesssim F_{\rm X,crit}$ and $F_{\rm 3-9 keV}\gtrsim F_{\rm X,crit}$ respectively. We find that the correlation slopes $b_1=0.61\pm0.03$ for GX 339--4 and $b_1=0.24\pm0.10$ for H 1743--322, which are much smaller than the second correlation slopes of $b_2\sim1.1-1.4$. For Swift J1753.5--0127, only the solid points with positive $\Gamma-F_{\rm 3-9 keV}$ correlation are
   fitted (solid line in bottom panel of Figure 3), and $b_2=1.10\pm0.16$. Due to insufficient data, the radio--X-ray correlation for XTE J1752--223 is not fitted with our data. In stead, we select the quasi-simultaneous radio/X-ray data points from \citet{br13} where the X-ray data are reduced from several different X-ray satellites (e.g., {\it Swift, MAXI} and/or {\it RXTE}). By excluding the data points with  $F_{\rm 3-9keV}\lesssim F_{\rm X,crit}$, we fit the radio--X-ray correlation for 17 data points and find $b_2=1.39\pm0.19$ (dashed line in bottom panel of Figure 4), where the critical X-ray flux is derived from the X-ray spectral evolution in our work (top panel of Figure 4).

%---------------------------------------------Figure 5-----------------------------------------------------------------------------------------------------------------------------------------
\begin{figure}[h]
%\vspace{-20mm}
\epsscale{0.9}
\plotone{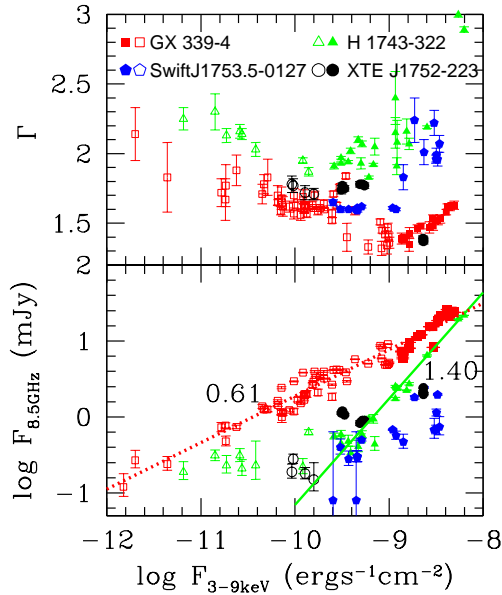}
%\vspace{-12mm}
\caption{The same as Figure 1, but for all sources. For comparison, the best fit of the data points with anti-correlated $\Gamma-F_{\rm 3-9keV}$ for GX 339-4 (dotted line) and the fit for the data points with  positive $\Gamma-F_{\rm 3-9keV}$ correlation for H 1743-322 (solid line) are also shown. \label{fig1}}
\end{figure}

   For comparison, we present the correlations of $\Gamma-F_{\rm X}$ and $F_{\rm R}-F_{\rm X}$ for four XRBs in Figure 5. The data points of GX 339--4 with anti-correlation of $\Gamma-F_{\rm X}$ stay in the former universal track (open squares, bottom panel). All solid points (including GX 339--4) with positive $\Gamma-F_{\rm X}$ correlation, roughly form the `outliers' track, even each source seems to follow its own trend. Some data points with anti-correlation of $\Gamma-F_{\rm X}$ (open circles and triangles) stay in the transition track.

\section{Conclusion and discussion}
  In this work, we analyze the X-ray data from {\it RXTE} for a sample of four XRBs with multiple, quasi-simultaneous radio and X-ray observation, and these sources either stay in the universal track or stay in the `outliers' track as reported in former works. We find that the radio--X-ray correlation is tightly correlated to the X-ray spectral evolution. The data points with positive $\Gamma-F_{\rm 3-9 keV}$ correlation normally stay in the steeper `outliers' track, while the data points with anti-correlation of $\Gamma-F_{\rm 3-9 keV}$ either stay in a shallower universal track or stay in a roughly flat transition track. Both $\Gamma-F_{\rm 3-9 keV}$ correlation and radio--X-ray correlation are changed at similar critical X-ray flux, which support that the different tracks of radio--X-ray correlation should be regulated by different accretion processes if the X-ray emission originate from accretion flows.

  The anti- and positive correlations between $\Gamma$ and $F_{\rm 3-9 keV}$, occurring below and above a critical flux, $F_{\rm X,crit}$, may be indicative for some `switch' in radiation mechanism at a certain critical accretion rate. We note that the harder X-ray spectral indices for 17 data points that fitted with a broken power-law are used in investigating of X-ray spectral evolution, which will not affect our main conclusion because both two spectral indices are positively correlated to the X-ray fluxes (see Table 3). Adopting a typical BH mass of $10\msun$ \citep[][and references therein]{ru13} and a distance of 8 kpc \citep[][and references therein]{corb13} for these XRBs, we find that the critical Eddington ratio for the transition of anti- and positive $\Gamma-F_{\rm 3-9 keV}$ correlation is $L_{\rm bol}/L_{\rm Edd}\sim$ 1\% for H 1743--322, Swift J1753.5--0127 and XTE J1752--223, and is $\sim$ 10\% for GX 339--4, where the bolometric luminosity, $L_{\rm bol}$, is estimated from power-law emission at 0.1--200 keV with the fitted X-ray photon index and flux. The critical Eddington ratio around several percent is roughly consistent with theoretical prediction for transition of ADAF and disk/corona \citep[e.g.,][]{na98}. In ADAF model, the optical depth will increase with increasing of accretion rate, increasing the Compton $y$-parameter, thereby leading to a harder X-ray spectrum, where electron temperature remain roughly unchanged with varying of accretion rate. In disk-corona model, the optical depth of corona will decrease when accretion rate increases due to stronger cooling, which lead to a smaller $y$-parameter and a softer X-ray spectrum. The detailed calculations of ADAF and disk-corona model do predict the anti- and positive correlations of photon index and Eddington ratio as observed in XRBs \citep[e.g.,][]{cao09,ql13}. Therefore, the different accretion modes may be the physical reason for the change of the X-ray spectral evolution. It should be noted that different values of viscosity parameter $\alpha$ will lead to different critical accretion rates for disk transition \citep[e.g.,][]{liu09}. For a larger $\alpha$, the critical accretion rate for the disk transition will also be large, which may explain why GX 339-4 has a higher critical flux for X-ray spectral evolution than other three XRBs. For XTE J1752-223, there are four data points that evidently deviate from the X-ray spectral evolution of other points (see top panel of Figure 4), which may be caused by the clearly different properties of accretion flow (e.g., magnetic field and or viscosity etc) compared that in other data points. The broadband data and more careful analysis are wished to further investigate their accretion properties, which will be our future work.

  The radio--X-ray correlation for the outliers of XRBs have been explored in several XRBs, where the correlation slopes are $\sim1.1-1.7$ for a certain XRB \citep[][]{sole10,ru10,cori11, br13}.  \citet{gall12} re-explored the radio--X-ray correlation for the `outliers' as a whole, and found that the correlation slope is $\sim0.98$, where the shallower relation may be caused by the data points in the `transition' track are also included in their fitting. From the X-ray spectral evolution, the `outliers' can well separated into two tracks: the `outliers' track with only positive correlation of $\Gamma-F_{\rm 3-9 keV}$ and the transition track with anti-correlation of $\Gamma-F_{\rm 3-9 keV}$. Excluding the data points in the transition track, the radio--X-ray correlation of the `outliers' track will become steeper, and this is the reason why our correlation slopes are larger than those reported in former works. It is interesting to note that the {\it bright} hard state of GX 339--4 with positive correlation of $\Gamma-F_{\rm 3-9 keV}$ also shows a much steeper radio--X-ray correlation ($\sim1.1$) than the universal one. Therefore, all these data points may also belong to the `outliers' track even this source was divided into the universal track in former works \citep[][]{cori11,corb13}. The correlation slopes are different for different `outliers' of XRBs, which may be caused by the different properties of accreting matter (e.g., magnetic field etc.). Due to the different trends of radio--X-ray correlation in each XRB, we don't fit the XRBs in the `outliers' track as a whole in $F_{\rm R}-F_{\rm X}$ relation. The luminosity--luminosity relation and/or Eddington-scaled luminosity--luminosity relation for a sample of XRBs may be more intrinsic, and, however, the distances and BH masses are still very uncertain for these several XRBs.

  \citet{cori11} proposed that different radio--X-ray correlations are possibly regulated by different accretion processes, which is supported by our results on the hard X-ray spectral evolution. For a classical jet, the jet power, $P_{\rm jet}$, is a fraction of accretion power and $P_{\rm jet}\propto \dot{M}$ \citep[e.g.,][]{falc95,hein03}. In the optically thick jet, the radio emission and jet power follow a scaling of $L_{\rm R}\propto P_{\rm jet}^{\sim 1.4}$ \citep[e.g.,][]{hein03,cori11}. The radiatively inefficient ADAF is expected to produce X-ray emission with $L_{\rm X}\propto \dot{M}^{q}$ and $q\sim 2.0$ (e.g., \citealt{merl03,yc05,wu06}). Therefore, the ADAF model can not only explain the anti-correlation of $\Gamma-F_{\rm 3-9 keV}$ \citep[e.g.,][]{ql13}, but can also  reproduce the observed universal correlation of $L_{\rm R}\propto L_{\rm X}^{\sim 0.7}$. For a magnetically heated corona above and below SSD, $q\simeq1$ for a gas pressure dominated SSD, while $q<1$ for a radiation pressure dominated SSD \citep[e.g.,][]{mf02}. We expect $L_{\rm R}\propto L_{\rm X}^{\sim 1.4}$, which is supported by the detailed calculations based on the jet formation from the disk-corona system \citep[][]{hww14}. Therefore, the disk-corona model can explain both the steeper radio--X-ray correlation and the positive $\Gamma-F_{\rm 3-9 keV}$ relation simultaneously. From our results, the dual tracks of radio--X-ray correlation are most possibly regulated by different accretion processes. The `radio quiet outliers' compared those sources in the universal track are caused by `X-ray loud' due to the increasing of the radiative efficiency as change of accretion mode. The transition track may be regulated by a transition state of accretion flow. The radiative efficiency will increase very fast at a critical accretion rate (e.g., from low radiative efficiency of ADAF to high radiative efficiency of disk-corona). If this is the case, we expect the X-ray luminosity will increase much faster than the radio luminosity if the accretion rate approach to the critical rate for disk transition, which can roughly explain the flat transition track. However, the physical reason for disk transition is still unclear, and the possible models include truncated SSD-ADAF \citep[e.g.,][]{lu04}, clumpy-ADAF \citep[e.g.,][]{wang12}, magnetically dominated accretion flow \citep[e.g.,][]{oda12} and luminous hot accretion flow \citep[e.g.,][]{yu01,xy12}, etc. These transitional disks are more or less ADAF-like, so the anti-correlated $\Gamma-F_{\rm 3-9 keV}$ will be also expected, and the detailed spectral calculation is beyond of this work.

\section*{Acknowledgements}
 We thanks Xinwu Cao and members of HUST astrophysics group for many useful discussions and comments. This work is supported by the NSFC (grants 11303010, 11103003 and 11133005) and New Century Excellent Talents in University (NCET-13-0238).

%-------------------------------------------------Table 1-----------------------------------------------------------------------------------------------------------------------------------
\clearpage

\scriptsize
\begin{longtable}{ccccccc}
\caption{Summary of data for GX 339--4.}\\
\hline
%\textbf{First entry} & \textbf{Second entry} & \textbf{Third entry} & \textbf{Fourth entry} \\
Date  & Obs. Id & $F_{\rm R}$   & $F_{\rm X}$              & $\Gamma$&   $\chi_{\nu}^2$(dof)& Model$^a$\\
     &         &  8.6 or 9.0 GHz&     3--9 keV            &         &           &      \\
     &         & mJy             &$10^{-11}\rm erg/s/cm^2$             &         &           &      \\
\hline
\endfirsthead
\multicolumn{4}{l}%
{\tablename\ \thetable\ -- \textit{Continued from previous page}} \\
\hline
%\textbf{First entry} & \textbf{Second entry} & \textbf{Third entry} & \textbf{Fourth entry}& \textbf{Second entry} & \textbf{Third entry} & \textbf{Fourth entry} \\
Date  & Obs. Id & $F_{\rm R}$   & $F_{\rm X}$              & $\Gamma$&   $\chi_{\nu}^2$(dof)& Model$^a$\\
     &         &  8.6 or 9.0 GHz&     3--9 keV            &         &           &      \\
     &         & mJy             &$10^{-11}\rm erg/s/cm^2$             &         &           &      \\
\hline
\endhead
\hline \multicolumn{4}{l}{\textit{Continued on next page}} \\
\endfoot
\hline
\endlastfoot
1997-02-03&20181-01-01-01&9.10$\pm$0.10&$96.70_{-0.30}^{+0.30}$&$1.47_{-0.03}^{+0.02}$&0.86(65)&diskbb+gau+pow\\
1997-02-10&20181-01-02-00&8.20$\pm$0.20&$84.26_{-0.24}^{+0.24}$&$1.45_{-0.02}^{+0.02}$&1.03(65)&diskbb+gau+pow\\
1997-02-17&20181-01-03-00&8.70$\pm$0.20&$80.68_{-0.23}^{+0.23}$&$1.51_{-0.02}^{+0.02}$&0.90(65)&diskbb+gau+pow\\
1999-03-03&40108-01-04-00&5.74$\pm$0.06&$43.32_{-0.16}^{+0.16}$&$1.59_{-0.03}^{+0.03}$&1.31(66)&diskbb+gau+pow\\
1999-04-22&40108-02-02-00&3.20$\pm$0.06&$20.44_{-0.08}^{+0.09}$&$1.60_{-0.01}^{+0.01}$&1.27(61)&gau+pow\\
1999-05-14&40108-02-03-00&1.44$\pm$0.06&$6.60_{-0.06}^{+0.06}$&$1.65_{-0.02}^{+0.02}$&0.89(63)&pow\\
1999-06-25&40105-02-02-00&0.24$\pm$0.05&$0.44_{-0.07}^{+0.07}$&$1.83_{-0.25}^{+0.26}$&0.49(63)&pow\\
1999-07-07&40108-03-01-00&0.12$\pm$0.04&$0.15_{-0.03}^{+0.03}$&$2.18_{-0.20}^{+0.21}$&0.50(63)&pow\\
1999-08-17&40108-03-03-00&0.27$\pm$0.07&$0.20_{-0.04}^{+0.04}$&$2.14_{-0.19}^{+0.20}$&0.58(63)&pow\\
2002-04-03&70109-01-02-00&5.95$\pm$0.15&$139.76_{-0.42}^{+0.42}$&$1.38_{-0.02}^{+0.02}$&0.90(59)&diskbb+gau+pow\\
2002-04-07&70109-01-01-00&8.27$\pm$0.07&$293.89_{-1.09}^{+1.09}$&$1.48_{-0.03}^{+0.02}$&1.02(59)&diskbb+gau+pow\\
2003-05-25&80102-04-07-00&0.77$\pm$0.06&$1.82_{-0.13}^{+0.13}$&$1.67_{-0.15}^{+0.15}$&0.61(64)&pow\\
2004-02-13&80102-04-57-01&1.13$\pm$0.08&$7.44_{-0.23}^{+0.23}$&$1.62_{-0.07}^{+0.07}$&0.72(64)&pow\\
2004-02-24&80102-04-58-01&1.84$\pm$0.20&$24.74_{-0.29}^{+0.29}$&$1.52_{-0.02}^{+0.02}$&0.77(64)&pow\\
2004-03-16&90118-01-05-00&4.88$\pm$0.06&$90.21_{-0.41}^{+0.41}$&$1.42_{-0.05}^{+0.03}$&0.97(60)&diskbb+gau+pow\\
2004-03-17&90118-01-06-00&4.84$\pm$0.11&$95.23_{-0.48}^{+0.50}$&$1.34_{-0.06}^{+0.05}$&0.66(60)&diskbb+gau+pow\\
2004-03-18&90118-01-07-00&4.98$\pm$0.11&$98.89_{-0.32}^{+0.32}$&$1.40_{-0.03}^{+0.02}$&1.14(60)&diskbb+gau+pow\\
2004-03-19&80132-01-15-00&5.20$\pm$0.10&$100.62_{-0.38}^{+0.38}$&$1.36_{-0.04}^{+0.04}$&0.85(60)&diskbb+gau+pow\\
2005-04-24&90704-01-13-01&4.23$\pm$0.08&$27.65_{-0.27}^{+0.27}$&$1.63_{-0.02}^{+0.02}$&0.98(64)&pow\\
2005-04-28&91095-08-06-00&3.46$\pm$0.13&$18.47_{-0.11}^{+0.11}$&$1.61_{-0.01}^{+0.01}$&0.99(62)&gau+pow\\
2005-04-29&91095-08-07-00&3.32$\pm$0.10&$16.49_{-0.10}^{+0.10}$&$1.61_{-0.01}^{+0.01}$&0.96(62)&gau+pow\\
2005-04-30&91095-08-08-00&2.94$\pm$0.07&$15.63_{-0.10}^{+0.10}$&$1.61_{-0.01}^{+0.01}$&1.17(64)&pow\\
2005-05-02&90165-01-01-02&1.92$\pm$0.14&$12.57_{-0.16}^{+0.16}$&$1.61_{-0.03}^{+0.03}$&1.07(64)&pow\\
2005-05-04&91105-04-18-00&1.99$\pm$0.10&$12.46_{-0.32}^{+0.32}$&$1.63_{-0.06}^{+0.06}$&0.80(64)&pow\\
2005-05-06&90704-01-14-00&1.69$\pm$0.12&$10.64_{-0.18}^{+0.18}$&$1.63_{-0.04}^{+0.04}$&0.60(64)&pow\\
2005-05-12&90704-01-14-02&1.00$\pm$0.18&$7.26_{-0.17}^{+0.17}$&$1.65_{-0.05}^{+0.05}$&1.00(64)&pow\\
2007-06-06&92704-03-25-00&2.63$\pm$0.18&$15.28_{-0.18}^{+0.18}$&$1.63_{-0.03}^{+0.03}$&1.04(64)&pow\\
2007-06-11&92704-03-29-00&2.01$\pm$0.15&$12.98_{-0.13}^{+0.13}$&$1.63_{-0.02}^{+0.02}$&1.31(64)&pow\\
2007-06-25&92704-03-38-00&1.69$\pm$0.05&$12.51_{-0.15}^{+0.15}$&$1.61_{-0.03}^{+0.03}$&1.03(64)&pow\\
2007-06-29&92704-03-42-00&2.00$\pm$0.20&$13.28_{-0.18}^{+0.18}$&$1.57_{-0.03}^{+0.03}$&0.92(64)&pow\\
2007-07-04&92704-03-44-01&2.10$\pm$0.20&$17.46_{-0.25}^{+0.25}$&$1.59_{-0.03}^{+0.03}$&0.79(64)&pow\\
2007-07-13&92704-03-47-00&2.66$\pm$0.05&$24.82_{-0.34}^{+0.34}$&$1.57_{-0.03}^{+0.03}$&0.68(64)&pow\\
2007-08-23&93409-01-05-03&2.95$\pm$0.07&$35.63_{-0.25}^{+0.25}$&$1.40_{-0.10}^{+0.08}$&0.95(60)&diskbb+gau+pow\\
2007-12-27&93409-01-20-00&0.48$\pm$0.07&$1.84_{-0.13}^{+0.13}$&$1.77_{-0.15}^{+0.16}$&0.61(64)&pow\\
2008-06-26&93076-08-01-00&1.16$\pm$0.10&$7.08_{-0.10}^{+0.10}$&$1.59_{-0.03}^{+0.03}$&0.97(64)&pow\\
2008-07-05&93076-08-03-00&1.24$\pm$0.07&$8.68_{-0.15}^{+0.15}$&$1.62_{-0.04}^{+0.04}$&0.70(64)&pow\\
2008-07-16&93076-08-05-05&1.51$\pm$0.06&$11.58_{-0.23}^{+0.23}$&$1.56_{-0.04}^{+0.04}$&0.93(64)&pow\\
2008-08-18&93108-01-01-02&1.18$\pm$0.10&$9.72_{-0.16}^{+0.16}$&$1.57_{-0.04}^{+0.04}$&0.67(64)&pow\\
2008-10-10&93702-04-03-00&0.73$\pm$0.10&$1.65_{-0.08}^{+0.08}$&$1.72_{-0.10}^{+0.10}$&0.49(64)&pow\\
2010-01-21&95409-01-02-02&5.05$\pm$0.05&$59.93_{-0.29}^{+0.29}$&$1.33_{-0.05}^{+0.05}$&0.58(60)&diskbb+gau+pow\\
2010-02-12&95409-01-06-00&5.90$\pm$0.10&$82.83_{-0.37}^{+0.37}$&$1.32_{-0.05}^{+0.04}$&0.56(60)&diskbb+gau+pow\\
2010-03-02&95409-01-08-02&7.30$\pm$0.10&$134.94_{-0.57}^{+0.60}$&$1.39_{-0.04}^{+0.03}$&1.12(60)&diskbb+gau+pow\\
2010-03-04&95409-01-08-03&7.30$\pm$0.10&$145.24_{-0.51}^{+0.51}$&$1.39_{-0.03}^{+0.02}$&0.81(60)&diskbb+gau+pow\\
2010-03-06&95409-01-09-06&9.60$\pm$0.05&$163.27_{-0.74}^{+0.74}$&$1.35_{-0.05}^{+0.05}$&0.88(60)&diskbb+gau+pow\\
2010-03-07&95409-01-09-01&8.05$\pm$0.10&$164.11_{-0.86}^{+0.86}$&$1.43_{-0.04}^{+0.04}$&0.86(60)&diskbb+gau+pow\\
2010-03-14&95409-01-10-02&11.32$\pm$0.10&$199.22_{-0.91}^{+0.91}$&$1.46_{-0.03}^{+0.03}$&0.81(60)&diskbb+gau+pow\\
2010-03-16&95409-01-10-04&12.04$\pm$0.10&$218.29_{-0.69}^{+0.70}$&$1.48_{-0.02}^{+0.02}$&0.79(60)&diskbb+gau+pow\\
2010-03-20&95409-01-11-01&15.45$\pm$0.06&$268.48_{-1.14}^{+1.15}$&$1.47_{-0.04}^{+0.03}$&0.69(60)&diskbb+gau+pow\\
2010-03-22&95409-01-11-02&15.45$\pm$0.06&$291.93_{-1.38}^{+1.38}$&$1.53_{-0.03}^{+0.03}$&0.80(60)&diskbb+gau+pow\\
2010-03-24&95409-01-11-03&18.59$\pm$0.05&$329.61_{-1.37}^{+1.38}$&$1.52_{-0.03}^{+0.03}$&1.00(60)&diskbb+gau+pow\\
2010-03-26&95409-01-12-00&21.88$\pm$0.10&$352.09_{-1.19}^{+1.20}$&$1.53_{-0.03}^{+0.02}$&0.89(60)&diskbb+gau+pow\\
2010-03-30&95409-01-12-02&21.88$\pm$0.10&$405.22_{-1.48}^{+1.51}$&$1.58_{-0.02}^{+0.02}$&1.36(60)&diskbb+gau+pow\\
2010-03-31&95409-01-12-04&25.94$\pm$0.05&$414.09_{-1.40}^{+1.41}$&$1.62_{-0.02}^{+0.02}$&0.90(60)&diskbb+gau+pow\\
2010-04-03&95409-01-13-00&21.11$\pm$0.15&$442.55_{-1.38}^{+1.40}$&$1.61_{-0.02}^{+0.02}$&1.14(60)&diskbb+gau+pow\\
2010-04-05&95409-01-13-02&24.69$\pm$0.05&$467.72_{-1.45}^{+1.46}$&$1.63_{-0.02}^{+0.02}$&1.23(60)&diskbb+gau+pow\\
2010-04-06&95409-01-13-05&23.90$\pm$0.06&$491.54_{-1.91}^{+1.92}$&$1.63_{-0.03}^{+0.02}$&1.02(60)&diskbb+gau+pow\\
2011-02-12&96409-01-07-03&4.17$\pm$0.05&$29.68_{-0.31}^{+0.31}$&$1.76_{-0.02}^{+0.02}$&1.04(64)&pow\\
2011-02-14&96409-01-07-01&4.17$\pm$0.05&$25.68_{-0.27}^{+0.27}$&$1.71_{-0.02}^{+0.02}$&1.23(64)&pow\\
2011-02-15&96409-01-07-02&3.87$\pm$0.05&$34.65_{-0.30}^{+0.30}$&$1.84_{-0.02}^{+0.02}$&1.27(64)&pow\\
2011-02-17&96409-01-07-04&3.98$\pm$0.10&$20.19_{-0.24}^{+0.24}$&$1.64_{-0.03}^{+0.03}$&0.63(64)&pow\\
2011-02-19&96409-01-08-00&3.98$\pm$0.10&$16.14_{-0.21}^{+0.21}$&$1.70_{-0.03}^{+0.03}$&0.95(64)&pow\\
2011-02-21&96409-01-08-02&3.84$\pm$0.05&$12.16_{-0.27}^{+0.27}$&$1.69_{-0.05}^{+0.05}$&0.80(64)&pow\\
2011-02-24&96409-01-08-03&2.95$\pm$0.05&$9.72_{-0.23}^{+0.23}$&$1.70_{-0.05}^{+0.06}$&1.00(64)&pow\\
2011-02-28&96409-01-09-01&2.42$\pm$0.08&$7.20_{-0.25}^{+0.25}$&$1.68_{-0.08}^{+0.08}$&0.45(64)&pow\\
2011-03-02&96409-01-09-02&1.64$\pm$0.05&$6.55_{-0.21}^{+0.21}$&$1.70_{-0.07}^{+0.07}$&0.64(64)&pow\\
2011-03-06&96409-01-10-01&1.26$\pm$0.10&$5.11_{-0.28}^{+0.28}$&$1.83_{-0.13}^{+0.13}$&0.62(64)&pow\\
2011-03-08&96409-01-10-02&1.26$\pm$0.10&$4.69_{-0.14}^{+0.14}$&$1.78_{-0.07}^{+0.07}$&0.67(64)&pow\\
2011-03-10&96409-01-10-03&1.38$\pm$0.08&$4.47_{-0.14}^{+0.14}$&$1.71_{-0.07}^{+0.07}$&0.52(64)&pow\\
2011-03-19&96409-01-12-01&0.74$\pm$0.04&$2.38_{-0.12}^{+0.12}$&$1.88_{-0.11}^{+0.12}$&0.64(64)&pow\\
\hline
\end{longtable}
\vspace{-4mm}
\begin{minipage}{95mm}
\centering
$^a$  an absorption model, {\it phabs}, was used in all fittings.
\end{minipage}

%-----------------------------------------------------------Table 2-------------------------------------------------------------------------------------------------------
\begin{table} [ht]
\vspace{4mm}
\begin{minipage}{150mm}
\centerline{TABLE 2. Summary of data for H 1743-322.}
\end{minipage}
\centering
%\tiny
\scriptsize
\begin{tabular}{cccccccc}
\hline
Date  &Obs. Id& $F_{\rm R}$&$F_{\rm X}$              & $\Gamma$ & $\Gamma_{\rm H}$&   $\chi^2$& Model$^a$ \\
    &          &  8.5 GHz  & 3--9 keV                   &         &          &   &   \\
     &         & mJy        &$10^{-11}\rm erg/s/cm^2$&         &      &     &      \\
\hline
2003-04-01&80138-01-02-00&6.45 $\pm$0.12&$253_{-0.42}^{+0.41}$   &$1.56_{-0.01}^{+0.01}$&$2.19_{-0.01}^{+0.01}$   &1.70(88)&diskbb+gau+bknpow\\
2003-04-04&80138-01-03-00&21.81$\pm$0.13&$626_{-0.63}^{+0.63}$   &$2.27_{-0.01}^{+0.01}$&$2.89_{-0.02}^{+0.02}$   &1.08(88)&diskbb+gau+bknpow\\
2003-04-06&80138-01-04-00&19.43$\pm$0.16&$545_{-0.55}^{+0.55}$   &$2.23_{-0.04}^{+0.04}$&$3.0_{-0.02}^{+0.02}$    &0.93(87)&diskbb+gau+bknpow\\
2003-11-05&80137-01-32-00&0.22 $\pm$0.05&$12.09_{-0.12}^{+0.11}$ &...                   &$1.95_{-0.02}^{+0.02}$   &0.96(88)&gau+pow\\
2004-11-01&90115-01-05-03&0.31 $\pm$0.06&$2.59_{-0.13}^{+0.13}$  &...                   &$2.16_{-0.05}^{+0.05}$   &0.83(88)&gau+pow\\
2008-01-29&93427-01-03-03&0.44 $\pm$0.09&$70.35_{-0.25}^{+0.25}$ &...                   &$2.05_{-0.06}^{+0.06}$   &1.05(87)&diskbb+gau+pow\\
2008-02-02&93427-01-04-00&0.52 $\pm$0.07&$50.19_{-0.19}^{+0.19}$ &...                   &$1.93_{-0.04}^{+0.04}$   &0.81(87)&diskbb+gau+pow\\
2008-02-04&93427-01-04-02&0.48 $\pm$0.08&$36.85_{-0.16}^{+0.16}$ &...                   &$1.94_{-0.01}^{+0.01}$   &1.23(89)&gau+pow\\
2008-02-06&93427-01-04-03&0.45 $\pm$0.09&$30.79_{-0.14}^{+0.15}$ &...                   &$1.93_{-0.01}^{+0.01}$   &1.00(89)&gau+pow\\
2008-02-08&93427-01-05-00&0.56 $\pm$0.05&$25.94_{-0.18}^{+0.18}$ &...                   &$1.91_{-0.02}^{+0.01}$   &1.01(89)&gau+pow\\
2008-02-18&93427-01-06-02&0.23 $\pm$0.12&$3.82_{-0.18}^{+0.18}$  &...                   &$2.03_{-0.05}^{+0.05}$   &0.82(89)&gau+pow\\
2008-02-20&93427-01-06-03&0.21 $\pm$0.05&$2.72_{-0.11}^{+0.11}$  &...                   &$2.14_{-0.04}^{+0.04}$   &0.79(89)&gau+pow\\
2008-02-22&93427-01-07-00&0.23 $\pm$0.06&$1.87_{-0.14}^{+0.14}$  &...                   &$2.13_{-0.05}^{+0.05}$   &0.92(89)&gau+pow\\
2008-02-24&93427-01-07-02&0.31 $\pm$0.05&$1.41_{-0.26}^{+0.26}$  &...                   &$2.30_{-0.13}^{+0.15}$   &1.04(89)&gau+pow\\
2008-10-07&93427-01-09-01&1.74 $\pm$0.07&$116.60_{-0.36}^{+0.36}$&$1.50_{-0.04}^{+0.04}$&$2.40_{-0.19}^{+0.27}$   &1.18(85)&diskbb+gau+bknpow\\
2008-10-08&93427-01-09-03&2.54 $\pm$0.08&$116.62_{-0.36}^{+0.36}$&$1.50_{-0.05}^{+0.04}$&$2.15_{-0.11}^{+0.13}$   &0.72(85)&diskbb+gau+bknpow\\
2008-10-09&93427-01-09-02&2.43 $\pm$0.09&$120.42_{-0.47}^{+0.47}$&$1.34_{-0.10}^{+0.11}$&$1.91_{-0.06}^{+0.06}$   &0.94(85)&diskbb+gau+bknpow\\
2008-10-11&93427-01-10-00&2.38 $\pm$0.11&$121.92_{-0.63}^{+0.63}$&$1.50_{-0.06}^{+0.11}$&$2.08_{-0.16}^{+0.23}$   &1.16(85)&diskbb+gau+bknpow\\
2008-11-04&93427-01-13-06&0.94 $\pm$0.12&$68.22_{-0.29}^{+0.29}$ &...                   &$1.92_{-0.01}^{+0.01}$   &1.42(91)&pow\\
2008-11-09&93427-01-14-02&0.94 $\pm$0.08&$60.86_{-0.39}^{+0.39}$ &...                   &$1.83_{-0.01}^{+0.01}$   &1.38(91)&pow\\
2009-05-29&94413-01-02-00&2.24 $\pm$0.03&$153.29_{-0.56}^{+0.57}$&$1.51_{-0.06}^{+0.06}$&$2.16_{-0.09}^{+0.12}$   &1.15(85)&diskbb+gau+bknpow\\
2009-05-30&94413-01-02-01&2.73 $\pm$0.10&$163.60_{-0.68}^{+0.68}$&$1.43_{-0.01}^{+0.05}$&$2.07_{-0.04}^{+0.05}$   &1.29(85)&diskbb+gau+bknpow\\
2009-07-07&94413-01-07-01&0.59 $\pm$0.06&$51.56_{-0.36}^{+0.36}$ &...                   &$2.07_{-0.02}^{+0.02}$   &0.94(91)&pow\\
2009-07-09&94413-01-07-02&0.41 $\pm$0.07&$46.94_{-0.36}^{+0.36}$ &...                   &$2.10_{-0.02}^{+0.02}$   &1.27(91)&pow\\
2009-07-11&94413-01-08-02&0.34 $\pm$0.06&$38.85_{-0.41}^{+0.41}$ &...                   &$2.02_{-0.02}^{+0.02}$   &0.89(91)&pow\\
2009-07-13&94413-01-08-00&0.59 $\pm$0.07&$30.82_{-0.21}^{+0.21}$ &...                   &$1.97_{-0.01}^{+0.01}$   &1.09(89)&gau+pow\\
2009-07-19&94413-01-09-00&0.63 $\pm$0.05&$13.95_{-0.22}^{+0.22}$ &...                   &$1.87_{-0.03}^{+0.03}$   &1.02(89)&gau+pow\\
2009-08-06&94413-01-11-02&0.19 $\pm$0.05&$0.65_{-0.15}^{+0.15}$  &...                   &$2.25_{-0.08}^{+0.08}$   &0.88(89)&gau+pow\\
\hline
\end{tabular}
%\caption{The list of observation ID  of {\it RXTE}, radio flux, unabsorbed X-ray flux and the spectral index of H 1743-322}
\begin{minipage}{150mm}
$^a$  an absorption model, {\it phabs}, was used in all fittings.
\end{minipage}
\end{table}

%--------------------------------------------Table 3----------------------------------------------------------------------------------------------------------
\begin{table} [t]
\begin{minipage}{150mm}
%\centering
\centerline{TABLE 3. Summary of data for Swift J1753.5-0127.}
\end{minipage}
\centering
%\tiny
\scriptsize
\begin{tabular}{cccccccc}
\hline
Date  &Obs. Id& $F_{\rm R}$&$F_{\rm X}$              & $\Gamma$ & $\Gamma_{\rm H}$&   $\chi^2$& Model$^a$ \\
    &          &  8.4 GHz  & 3--9 keV                   &         &           &   &   \\
     &         & mJy        &$10^{-11}\rm erg\rm /s/cm^2$&         &      &     &      \\
\hline
2005-07-04&91094-01-01-01&0.68$\pm$0.20&$300.44_{-0.68}^{+0.67}$&$1.69_{-0.02}^{+0.02}$&$2.22_{-0.09}^{+0.11}$ &1.15(84)&diskbb+gau+bknpow\\
2005-07-06&91423-01-01-04&0.74$\pm$0.20&$342.78_{-0.74}^{+0.75}$&$1.71_{-0.02}^{+0.02}$&$2.07_{-0.06}^{+0.08}$ &1.12(84)&diskbb+gau+bknpow\\
2005-07-07&91423-01-01-00&0.68$\pm$0.68&$315.09_{-0.82}^{+0.83}$&$1.64_{-0.01}^{+0.01}$&$1.96_{-0.03}^{+0.04}$ &1.15(84)&diskbb+gau+bknpow\\
2005-07-08&91094-01-02-01&1.96$\pm$0.04&$327.10_{-0.74}^{+0.75}$&$1.66_{-0.02}^{+0.02}$&$1.96_{-0.05}^{+0.07}$ &1.03(84)&diskbb+gau+bknpow\\
2005-07-10&91094-01-02-02&1.14$\pm$0.12&$318.24_{-0.70}^{+0.70}$&$1.63_{-0.02}^{+0.02}$&$1.99_{-0.05}^{+0.06}$ &0.86(84)&diskbb+gau+bknpow\\
2005-07-19&91423-01-03-04&2.42$\pm$0.05&$234.77_{-0.56}^{+0.56}$&$1.59_{-0.02}^{+0.02}$&$2.01_{-0.09}^{+0.13}$ &0.95(84)&diskbb+gau+bknpow\\
2005-07-26&91423-01-04-04&1.81$\pm$0.09&$186.15_{-0.47}^{+0.47}$&$1.56_{-0.02}^{+0.02}$&$2.24_{-0.16}^{+0.19}$ &1.13(85)&diskbb+gau+bknpow\\
2005-08-03&91423-01-05-02&0.47$\pm$0.10&$141.72_{-0.39}^{+0.39}$&$1.55_{-0.02}^{+0.02}$&$1.83_{-0.09}^{+0.15}$ &1.01(85)&diskbb+gau+bknpow\\
2005-08-07&91423-01-06-01&0.57$\pm$0.07&$118.55_{-0.36}^{+0.36}$&...&$1.60_{-0.01}^{+0.01}$                    &1.22(91)&pow\\
2005-08-11&91423-01-06-03&0.68$\pm$0.01&$109.05_{-0.34}^{+0.33}$&...&$1.61_{-0.01}^{+0.01}$                    &1.61(91)&pow\\
2005-10-22&91423-01-17-00&0.28$\pm$0.06&$37.03_{-0.22}^{+0.22}$ &...&$1.60_{-0.01}^{+0.01}$                    &1.17(91)&pow\\
2005-11-19&91423-01-21-00&0.40$\pm$0.09&$30.40_{-0.22}^{+0.22}$ &...&$1.60_{-0.02}^{+0.02}$                    &0.78(91)&pow\\
2006-03-11&92404-01-02-00&0.08$\pm$0.07&$25.33_{-0.22}^{+0.22}$ &...&$1.65_{-0.02}^{+0.02}$                    &0.90(91)&pow\\
2007-07-01&93105-01-08-00&0.30$\pm$0.05&$45.52_{-0.20}^{+0.20}$ &...&$1.60_{-0.01}^{+0.01}$                    &1.09(91)&pow\\
2007-07-08&93105-01-09-00&0.30$\pm$0.05&$45.89_{-0.20}^{+0.20}$ &...&$1.61_{-0.01}^{+0.01}$                    &0.82(91)&pow\\
2007-07-15&93105-01-10-00&0.30$\pm$0.05&$46.01_{-0.20}^{+0.20}$ &...&$1.59_{-0.01}^{+0.01}$                    &1.03(91)&pow\\
2007-07-22&93105-01-11-00&0.08$\pm$0.08&$44.70_{-0.26}^{+0.25}$ &...&$1.60_{-0.01}^{+0.01}$                    &0.83(91)&pow\\
2009-06-09&93105-02-33-00&0.50$\pm$0.02&$50.95_{-0.22}^{+0.21}$ &...&$1.62_{-0.01}^{+0.01}$                    &1.09(91)&pow\\
\hline
\end{tabular}
\begin{minipage}{150mm}
$^a$  an absorption model, {\it phabs}, was used in all fittings.
\end{minipage}

\end{table}

%--------------------------------------------------Table 4-----------------------------------------------------------------------------------------------------

\begin{table} [t]
\vspace{-85mm}
\begin{minipage}{150mm}
\centerline{TABLE 4. Summary of data for XTE J1752-223.}
\end{minipage}
%\tiny
\centering
\scriptsize
\begin{tabular}{lccccccc}
\hline
Date  &Obs. Id& $F_{\rm R}$&$F_{\rm X}$              & $\Gamma$ & $\Gamma_{\rm H}$&   $\chi^2$& Model$^a$ \\
    &          &  8.4 GHz  & 3--9 keV                   &         &           &   &   \\
     &         & mJy        &$10^{-11}\rm erg\rm /s/cm^2$&         &      &     &      \\
\hline
2009-10-30&94331-01-02-01&2.42$\pm$0.05&$231.08_{-0.68}^{+0.68}$&$1.38_{-0.02}^{+0.01}$&0.91(60)&diskbb+gau+pow\\
2009-11-01&94331-01-02-05&1.99$\pm$0.08&$231.37_{-0.69}^{+0.69}$&$1.37_{-0.02}^{+0.02}$&0.61(60)&diskbb+gau+pow\\
2009-11-05&94331-01-02-17&2.05$\pm$0.07&$232.15_{-0.86}^{+0.87}$&$1.38_{-0.02}^{+0.02}$&0.93(60)&diskbb+gau+pow\\
2009-11-08&94331-01-03-05&2.10$\pm$0.10&$228.82_{-0.70}^{+0.71}$&$1.39_{-0.02}^{+0.02}$&0.86(60)&diskbb+gau+pow\\
2010-04-13&95360-01-12-03&0.90$\pm$0.05&$54.57_{-0.41}^{+0.41}$&$1.77_{-0.02}^{+0.02}$&1.25(64)&pow\\
2010-04-15&95360-01-12-04&0.90$\pm$0.05&$52.05_{-0.49}^{+0.49}$&$1.78_{-0.02}^{+0.02}$&1.23(64)&pow\\
2010-04-17&95702-01-01-01&0.83$\pm$0.06&$48.92_{-0.39}^{+0.39}$&$1.78_{-0.02}^{+0.02}$&1.23(64)&pow\\
2010-04-27&95702-01-02-04&1.07$\pm$0.04&$33.59_{-0.32}^{+0.32}$&$1.75_{-0.02}^{+0.02}$&0.84(64)&pow\\
2010-04-28&95702-01-02-05&1.20$\pm$0.03&$31.88_{-0.31}^{+0.31}$&$1.77_{-0.02}^{+0.02}$&1.08(64)&pow\\
2010-04-29&95702-01-02-06&1.13$\pm$0.04&$30.95_{-0.41}^{+0.41}$&$1.74_{-0.03}^{+0.03}$&1.06(64)&pow\\
2010-05-17&95702-01-05-03&0.28$\pm$0.04&$9.59_{-0.19}^{+0.19}$&$1.77_{-0.05}^{+0.05}$&0.83(64)&pow\\
2010-05-18&95702-01-05-04&0.19$\pm$0.04&$9.26_{-0.24}^{+0.24}$&$1.78_{-0.06}^{+0.06}$&0.84(64)&pow\\
2010-06-03&95702-01-07-03&0.18$\pm$0.03&$12.78_{-0.28}^{+0.28}$&$1.72_{-0.05}^{+0.05}$&0.82(64)&pow\\
2010-06-11&95702-01-09-00&0.15$\pm$0.06&$15.81_{-0.31}^{+0.31}$&$1.71_{-0.04}^{+0.04}$&0.77(64)&pow\\
\hline
\end{tabular}
\begin{minipage}{140mm}
$^a$  an absorption model, {\it phabs}, was used in all fittings.
\end{minipage}
\end{table}

%----------------------------------------Table 5----------------------------------------------------------------------------------------------------
\begin{table}  [h]
\vspace{-90mm}
\begin{minipage}{100mm}
\centerline{TABLE 5. Summary of the radio--X--ray fits.}
\end{minipage}
\centering
%\tiny
\scriptsize
\begin{tabular}{ll|cc|cc}
\hline
Source              &   $F_{\rm crit,X}$$^a$ &  $k_1$$^b$         & $b_1$              &  $k_2$$^c$          &  $b_2$         \\
\hline
GX 339--4           &   $1.3^{+0.7}_{-0.5}\times10^{-9}$    & 6.39$\pm$0.30  &  0.61$\pm$0.03     & 10.93$\pm$2.39  & 1.14$\pm$0.27   \\
H1743--322          &   $2.5^{+0.6}_{-0.5}\times10^{-10}$   & 1.91$\pm$1.04  &  0.24$\pm$0.10     & 12.83$\pm$1.29  & 1.40$\pm$0.14     \\
Swift J1753.5--0127 &   $\lesssim2.5\times10^{-10}$         &     \nodata    &  \nodata           & 9.55$\pm$1.23   &  1.10$\pm$0.13      \\
XTE J1752--223      &   $1.6\times^{+0.2}_{-0.3}10^{-10}$   &     \nodata    &  \nodata           & 12.83$\pm$1.39  &  1.39$\pm$0.19       \\
\hline
\end{tabular}

\begin{minipage}{115mm}
$^a$ The critical flux for the anti- and positive correlations of $\Gamma-F_{\rm 3-9 keV}$, and $1-\sigma$ errors are also included.\\
$^{b,c}$ The best fits for the data points with anti- and positive correlations of $\Gamma-F_{\rm 3-9 keV}$ with a function of $F_{\rm R}=kF_{\rm X}^b$ respectively.
\end{minipage}
\end{table}

\end{document}